\documentclass[a4paper,11pt]{article}
\pdfoutput=1 

\usepackage{jcappub} 

\usepackage{array}

\usepackage[T1]{fontenc} 
\usepackage{multirow}
\graphicspath{{figures/}}
\title{\boldmath Charge ratio of cosmic ray muons in momentum range $\sim$\,1 to 3\,GeV/c}

\newcolumntype{P}[1]{>{\centering\arraybackslash}p{#1}}

\author[a,b,1]{Raj Shah,\note{Corresponding author.}}
\author[a,b]{J. M. John,}
\author[c]{Suryanarayan Mondal,}
\author[d]{S. Pethuraj,}
\author[b]{G. Majumder,}
\author[a,e]{P. Shukla}

\affiliation[a]{Homi Bhabha National Institute,\\Mumbai-400094, India}
\affiliation[b]{Tata Institute of Fundamental Research,\\Mumbai-400005, India}
\affiliation[c]{Universit$\grave{a}$ \& INFN di Pisa,\\ Pisa-56127, Italy}
\affiliation[d]{Dept. of Physics, Thiagarajar College of Engineering,\\Madurai 625015, India}
\affiliation[e]{Bhabha Atomic Research Centre,\\Mumbai-400085, India}

\emailAdd{raj.shah@tifr.res.in}

\abstract{This work presents the measurements of the cosmic muon charge ratio as a function of full azimuthal angle and momentum within the range of 0.8 to 3.0\,GeV/c, using the mini-ICAL detector. The detector, comprising 10 layers of RPCs, has collected cosmic muon data since August 2018  till recent time, at an altitude of 160\,m above sea level at the Inter-Institutional Center for High Energy Physics in Madurai, India $(9^\circ56'\,N, 78^\circ00'\,E)$. The muon charge identification is achieved through the use of a magnetic field of strength $1.4$\,T. The analysis shows that the cosmic muon charge ratio, $R_\mu = N_{\mu^+}/N_{\mu^-}$, ranges from 1.1 to 1.2 and has small dependency on the zenith angle. The charge ratio's dependence on momentum and azimuthal angle is thoroughly examined for a wide range of zenith angle upto $50^\circ$. These measurements are compared with the predictions from various combinations of different hadronic models in CORSIKA extensive air shower simulations.}

\begin{document}
\maketitle
\flushbottom

\section{Introduction}
\label{sec:intro}
The mini-ICAL is an 85\,ton magnetized calorimeter located at Inter-Institutional Centre for High Energy Physics (IICHEP), Madurai, and has Resistive Plate Chambers (RPCs) as tracking devices for the cosmic ray muons. 
The primary cosmic ray protons, upon reaching the top of the atmosphere, interact with atmospheric gases, leading to inelastic interactions that produce hadronic showers. These showers generate charged pions and kaons, which undergo decay processes, resulting in the production of muons and neutrinos. The muons that propagate down to sea level are subsequently measured in mini-ICAL. Obtaining measurements of the cosmic ray muon charge ratio is crucial for refining calculations of atmospheric neutrino fluxes and tuning the hadronic models in the CORSIKA \cite{Heck:1998vt}. These measurements hold significance for precise assessments of neutrino oscillations in atmospheric neutrino experiments and the computation of backgrounds for neutrino experiments.

The extensive air shower (EAS) simulations employ various hadronic interaction models to compute the cross sections and particle production multiplicity in inelastic processes. The discrepancies observed between the EAS simulation and observed data can serve as a basis for refining the shower development model. While exploring enhancements in the hadronic interaction model at high energy, the mini-ICAL faces limitations in momentum resolution at high momentum due to its limited number of layers, short lever-arm and poor position resolution in comparison to conventional tracking devices.

The Earth's geomagnetic field exerts an influence on cosmic rays. Given that the earth's magnetic field is oriented in a north-south direction, positive particles experience an eastward deflection as they approach the detector from the west, while negative particles are deflected westward as they approach from the east. This phenomenon leads to the east-west asymmetry of cosmic rays, which becomes more pronounced at lower muon momenta and larger zenith angles.

The charge ratio, $R_\mu = N_{\mu^+}/N_{\mu^-}$ can vary between 1.2 to 1.3 for muons with energies ($<$\,3\,GeV), for a rigidity cutoff of $\sim$0.4\,GV at Lynnn Lake location \cite{Haino:2004nq}. The charge ratio seems to decrease for larger rigidity cutoff. At Tsukuba location, the charge ratio in the same energy range varies from 1.1 to 1.2, where the rigidity cutoff is 11\,GV \cite{Haino:2004nq}. 
 The azimuthal dependence of muon charge ratio has been observed by WILLI experiment \cite{Brancus:2008zz}, with muons at low energies ($< $\,1\,GeV) having mean zenith angle $35^{\circ}$, where the geomagnetic cutoff is 5.6\,GV. The ratio varies between 0.8 to 1.4. 
The vertical geomagnetic rigidity cut-off for Madurai is $\sim$17.4\,GV, which is quite large compared to the rigidity cut-off for the locations of other similar experiments. The mini-ICAL has the capability to distinguish $\mu^+$ and $\mu^-$ at $\sim$ 0.8\,GeV/c to $\sim$ 3.0\,GeV/c and can measure incline muon events up to $65^{\circ}$. However the number of observed events at higher zenith angle is drastically reduced due to trigger acceptance and geometry, consequently, this measurement is restricted upto $50^\circ$.


The paper is organized as follows. In Section 2, the mini-ICAL detector and its relevant components for this analysis are described. An overview of the different CORSIKA hadronic models used in the event generation is provided in Section 3, along with a description of the Monte Carlo simulation of the mini-ICAL detector. The analysis of experimental data and simulation is detailed in Section 4, including the selection of events, reconstruction of the trajectory of muon, unfolding procedure, and response matrix. The systematic uncertainties are discussed in Section 5. In Section 6, the results are presented, focusing on the momentum and the azimuthal angle dependence of the cosmic muon charge ratio at the Earth's surface. Finally, Section 7 summarizes our findings.


\section{The mini-ICAL detector}
\label{sec:mICAL}
\begin{figure}[htbp]
\centering 
\includegraphics[width=.8\textwidth]{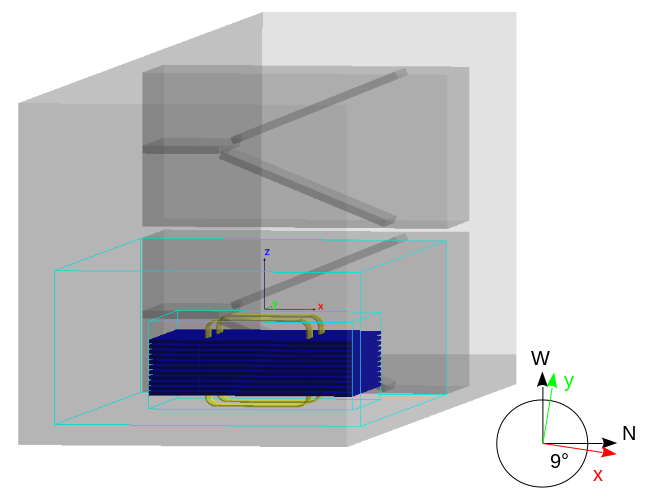}
\caption{\label{fig:miniICAL_in_building} The location of mini-ICAL in the building and it's orientation with respect to geo-magnetic axes.}
\end{figure}

The miniaturized ICAL detector, named mini-ICAL, was constructed at IICHEP Madurai for the purposes of prototyping the final ICAL detector and for the testing of the electronics in the fringe field of the magnet. Comprising 11 layers of 4\,m $\times$ 4\,m iron plates, each with a thickness of 5.6\,cm, the stack has a total weight of approximately 85\,tons.
The provision of spacing between the 11 layers allows for the placement of RPCs to track secondary cosmic particles.
The mini-ICAL is situated inside a building as shown in Figure\,\ref{fig:miniICAL_in_building}. The entire structure is inclined at a 9$^\circ$ angle relative to the geomagnetic north-south axis, towards the east.

\begin{figure}[htbp]
\centering 
\includegraphics[width=1.\textwidth]{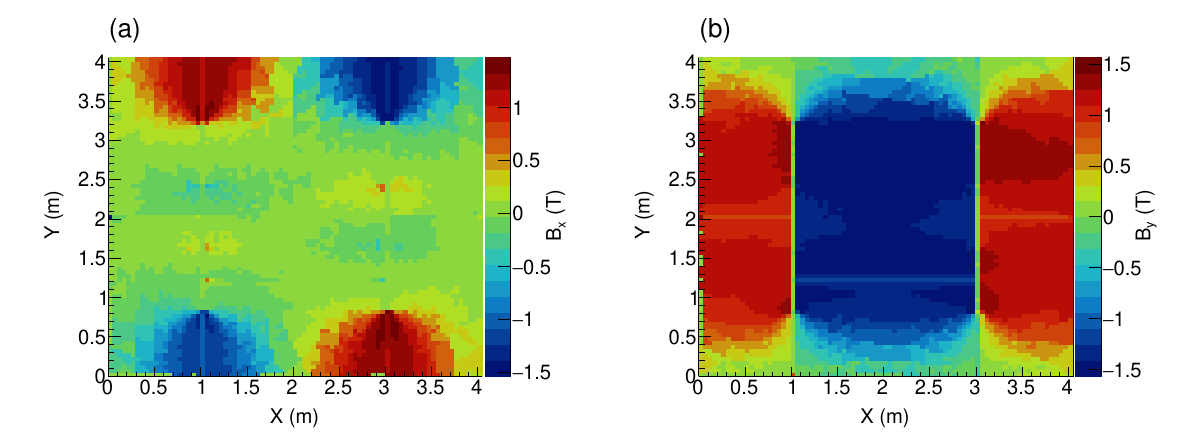}

\caption{The components of magnetic field in the detector co-ordinate: (a) $B_{x}$: The X component of the magnetic field in tesla. (b) $B_{y}$: The Y component of the magnetic field in tesla.}
\label{fig:magneticfieldmap}
\end{figure}

Configured as a toroidal magnet, the detector is designed to allocate a central region with dimensions of 2\,m $\times$ 4\,m, providing the capacity to house twenty 2\,m $\times$ 2\,m RPCs in this space. The coordinate system is defined such that the x-axis points towards the north, while the y-axis points towards the west. The vertical axis ($\theta = 0^\circ$) is designated as the z-axis. In the x-y plane, the azimuthal angle is measured from true geomagnetic north ($\phi=-9^\circ$) towards the west ($\phi = 81^\circ$). More details of the detector are given in \cite{John:2022fuy} and \cite{John:2023zqj}. For this study, only ten RPCs were utilized, positioned centrally within the mini-ICAL. They are numbered 0 to 9, from bottom to top. The magnetic field is in the negative Y direction as shown in Figure\,\ref{fig:magneticfieldmap}. The field deviations are less than 3.3\,$\%$ in the central region of the iron layers where RPCs are kept \cite{Khindri:2023fvf}.

A simulation and digitisation framework was developed for this detector setup along with some novice ideas as reported in \cite{John:2023zqj}.

\section{Monte Carlo Simulation}
\begin{figure}[htbp]
\centering
\includegraphics[width=0.5\textwidth]{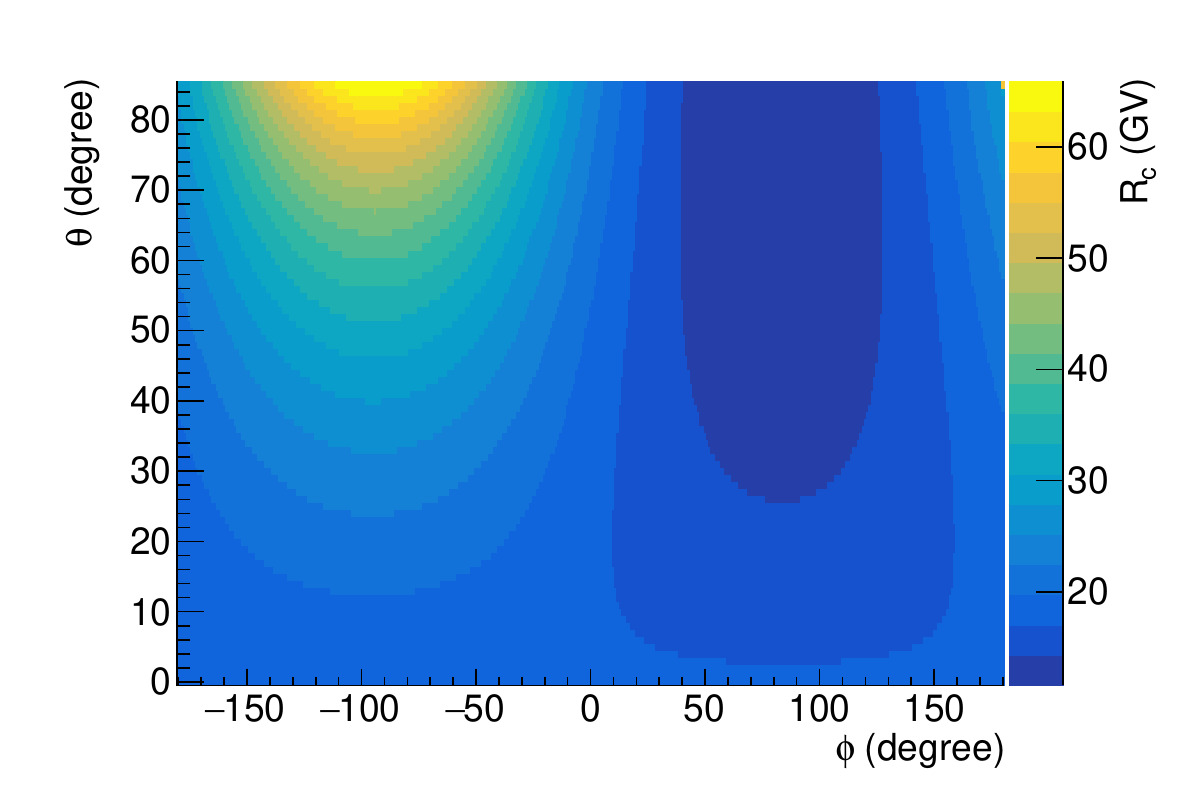}
\caption{Rigidity cutoff ($R_c$) for primary cosmic rays as a function of zenith angles ($\theta$) and azimuthal angles ($\phi$) at the Madurai location.}
\label{fig:rigiditycutoffmadurai}
\end{figure}

The detector simulation is done in two steps. The extensive air shower has been simulated with the CORSIKA(v7.6300) \cite{Heck:1998vt} package with the magnetic rigidity cut-off for the location of Madurai during the data taking period. The experiment being located close to the equator has quite larger rigidity cutoff values as compared to other similar experimental locations.  The primaries are generated within a range of 10\,GeV to 10$^6$\,GeV with a spectral index of $-$2.7. The rigidity cut-off is simulated using the International Geomagnetic Reference Field (IGRF), the 12th generation model by backtracking method~\footnote{The rigidity cutoff table for each $(\theta ,\phi )$ bins at Madurai location is calculated by Dr. Hariharan and Dr. P. K. Mohanty of TIFR and is obtained through private communication.} in all $(\theta ,\phi )$ ranges, where the zenith angle$(\theta)$ ranges from 0$^\circ$ - 85$^\circ$ and the azimuthal angle varies from $-$180$^\circ$ to +180$^\circ$. Figure \ref{fig:rigiditycutoffmadurai}, shows the rigidity cutoff ($R_c$) at the Madurai location as a function of direction. The secondary muons produced in the air shower with energy $\geq$ 100\,MeV are stored for the GEANT4 \cite{GEANT4:2002zbu} simulation. The ``CURVED'' atmosphere option is used to model the earth's atmosphere since the coverage of the zenith angle is upto 85$^\circ$.

  The earth's magnetic field values for the Madurai location is, $B_{x} = 40.431\,\mu T$ and $B_{z} = 4.705\,\mu T$ where, the $B_{x}$ is the horizontal component in x-direction towards North and $B_{z}$ is the vertical component downwards. For lower energy ranges, the hadron interaction models FLUctuating KAscade (FLUKA) \cite{Ferrari:2005zk}, Gamma Hadron Electron Interaction SHower code (GHEISHA) \cite{Fesefeldt:1985yw} and Ultra-relativistic Quantum Molecular Dynamics (UrQMD) \cite{Bleicher:1999xi} are used, and for high energy ranges, various hadron interaction models such as SIBYLL \cite{Fletcher:1994bd}, Quark Gluon String model with JETs  \cite{Kalmykov:1993qe} QGSJET 01C, QGSJETII 04 and Very Energetic NUclear Scattering (VENUS) \cite{Werner:1993uh} are used. All the possible combinations of these low and high energy models are simulated.
  
Particles produced by CORSIKA at the observation surface serve as input for the GEANT4 detector simulation.
The CORSIKA output includes the 4-vector of the muon.
The zenith angle, denoted as $\theta$, is determined by the angle between the momentum vector of the particle and the negative Z-axis, with the Z-axis upwards.

The X-, Y-, Z-axis has the same direction as that of the X-, Y-, Z-axis conventions in GEANT4. Since only the momentum along the Z-axis is denoted with negative sign, it has to be inverted to make it in the same convention as the detector coordinate system. Finally, the muon direction vector is rotated by $9^\circ$ w.r.t z-axis to match the orientation of the detector with the geomagnetic axis.

Muons generated at the surface are positioned above the top trigger RPC layer ($9^{th}$ layer) with directions provided by CORSIKA and then extrapolated to the bottom trigger layer ($6^{th}$ layer) after including the $9^{\circ}$ rotation in the $x-y$ plane. An event is simulated in GEANT4 only if the extrapolated trajectory intersects with the bottom trigger layer RPC including the uncertainty due to multiple scattering. During the simulation, the trajectory is extrapolated to the top of the building and GEANT4 takes care of the propagation towards the detector from that point. This approach effectively prevents the unnecessary storage of data which are not triggered due to geometrical constraints. The digitised information of energy deposited along the path of the charged particle in the sensitive detector, which in this case refers to the RPCs, is stored. During digitization, the parameters extracted from real detector data such as efficiency, dead strip information, multiplicity correlations, noise, etc. are reincorporated in the Monte Carlo sample to represent the real detector condition  \cite{John:2023zqj}. 

\section{Analysis of experimental data $\&$ simulation }
\label{sec:dataanalysis}
\begin{figure}[htbp]
\centering 
\includegraphics[width=.7\textwidth]{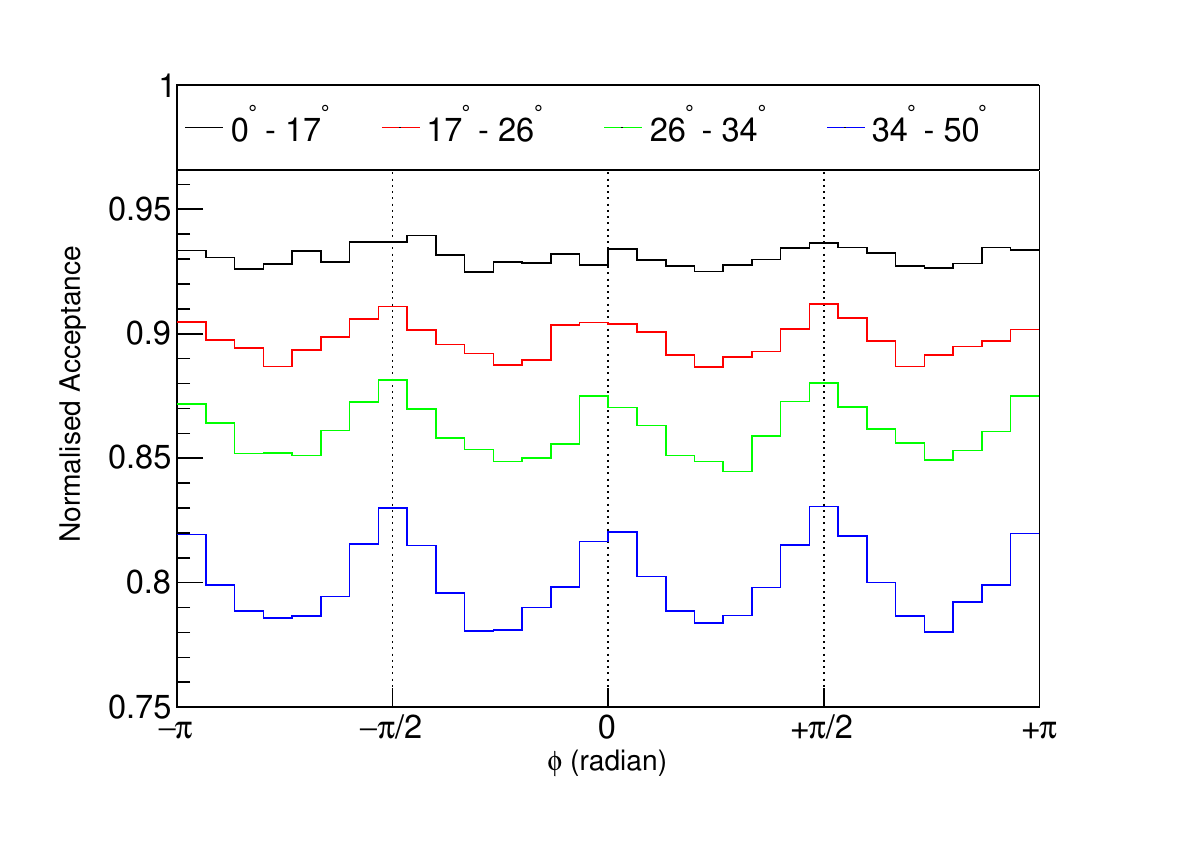}
\caption[Acceptance due to the geometry and trigger criteria.]{\label{fig:acceptance} Acceptance at different zenith angles due to the geometry and trigger criteria in detector co-ordinate. The efficiency of the trigger layers is not taken into consideration.}
\end{figure}

This paper showcases results derived from data collected between December 13, 2018, and December 30, 2018 having a trigger rate of $\sim$180\,Hz. Throughout this time-frame, the stack comprised merely 10 RPCs, strategically positioned within the central region where the magnetic field was very much uniform. The data with the magnetic field were taken during the day time. There were $\sim$70\,million events in the total sample. The data acquisition and trigger have been described in \cite{John:2023zqj}. The trigger requires the muons to pass through the top four RPC layers of the mini-ICAL, i.e., layer 6, 7, 8 and 9. The offline detector alignment algorithm is described in \cite{MAJUMDER201488}. The acceptance of the muons falling on the top trigger layer varies for different zenith angles, due to this trigger criteria and geometry as shown in Figure ~\ref{fig:acceptance}. The peaks correspond to North, West, East, South directions for larger zenith angle, which arises due to the geometrical acceptance of the detector.

The magnetic field in the central region of iron is along the negative Y direction as shown in Figure~\ref{fig:magneticfieldmap}. This is the default configuration of the magnetic field for mini-ICAL data taking. So the $\mu^+$ will be deflected towards the South and the $\mu^-$ will be deflected towards the North by the magnetic field in the centre of the mini-ICAL.

\subsection{Selection of events}
\label{subsec:selection}
Once the trigger is formed, the hit\footnote{In RPCs, the induced signal in the strip usually measures a few 100\,fC. Signals exceeding 100\,fC are regarded as hits or measurements, whereas the average noise pulses are around 10\,fC.} information is collected till $\sim$21\,$\mu$s after the trigger time. The long collection period is kept in order to study the possibility of determining the magnetic field within the iron from muon decays, given that it requires more time. In the muon flux measurement, the decayed hits were not needed, indeed it contains more noise hits. So the first step in the selection was to accept hits based on their time information. The average timing of muon hits within each layer is determined, and hits falling within $\pm$50\,ns of this average timing are retained for further analysis. The average time difference between the top layer (Layer 9) and all other layers is computed, serving as the basis for another selection criterion. For each event, the timing of hits in the top layer and the anticipated timing of hits in all other layers relative to the top layer are calculated based on the average time variance between layers. The hits within $\pm$10\,ns of this expected hit time in different layers are retained for further analysis. If the timing of hits in the top layer falls outside the initial range of $\pm$50\,ns, hits from the layer directly below (Layer 8) are considered and so on. This filtering process based on time is indirectly applied to the simulation. The noise is derived from the data subjected to the preceding timing criterion and is introduced into the simulation. Consequently, there's no requirement to impose a timing-based criterion in the simulated events.

In a specific Z plane, all potential combinations of X and Y strips within a $\pm$5\,ns window are combined to generate 2D-hits position. The time information of X- and Y-side strips are corrected for the distance traversed along the strips as well as along the electronic chains for each
  combinations, before imposing the $|\Delta t| \leq $5\,ns criterion and the timing information of the 2D-hits are taken from the average values of X- and Y-strip time. All nearby 2D-hits (upto 3 strips in x-side and y-side) are combined to form a cluster using the time criterion, $|\Delta t| \leq $5\,ns. The event selection process employs a track finder algorithm. 
A group of clusters meeting a track-like criterion across three out of five layers is combined and designated as a tracklet. This "three out of five" criterion is deliberately incorporated to accommodate inefficiencies. To determine which clusters form part of a track, an algorithm akin to a straight-line fit is employed, with a degree of flexibility to account for potential bending \cite{ICAL:2015stm}.
Events with only a single muon track are considered for the analysis.

\subsection{Reconstruction of muon tracks}
\label{subsec:reconstruction}
\begin{figure}[htbp]
\centering
\includegraphics[width=0.7\textwidth]{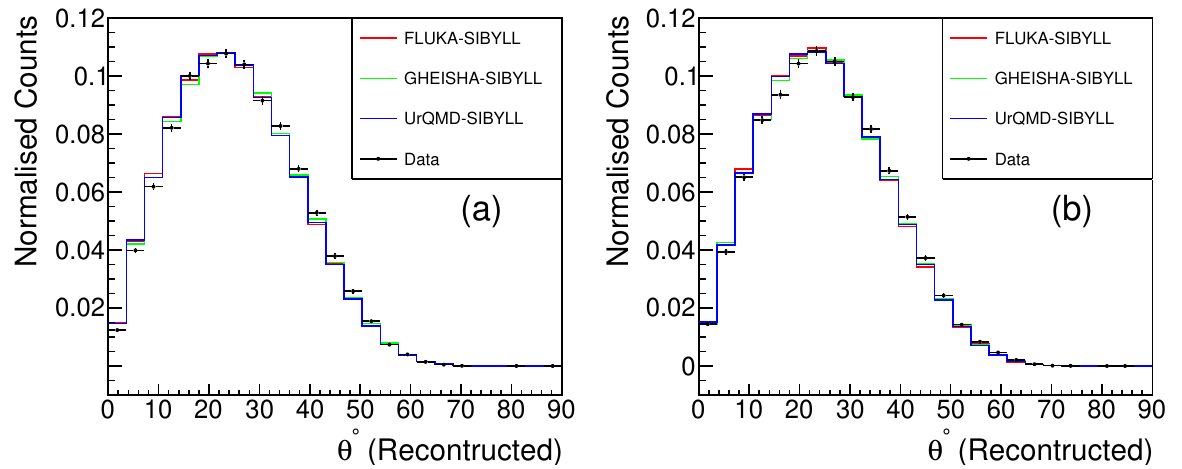}
\caption{Normalized reconstructed zenith angle distributions for (a) $\mu^{+}$ and (b) $\mu^{-}$.}
\label{theta_reco_distribution}
\end{figure}

The reconstruction of muons employs a Kalman filter algorithm \cite{Bhattacharya:2014tha}. This process involves track fitting to deduce information regarding the charge and momentum of the particle. The state vector is characterized by (x, y, dx/dz, dy/dz, q/p). Initially, the co-ordinate $x$ and $y$ are set from the valid signal in the top most layer and q/p for the first layer is set to zero, while dx/dz and dy/dz are determined using valid clusters in the top two layers. Subsequently, the state vector in the next layer is extrapolated by employing the equation of a charged particle in a magnetic field. Measurement errors arise due to finite strip width, inefficiency, multiple scattering, and the extrapolation error defined by a propagator matrix. This matrix encapsulates the overall error in each track parameter during extrapolation from point {\it l} to the subsequent point {\it l+dl}. Extrapolation variances originate from energy loss fluctuation and multiple scattering in the dense materials. Based on these errors, predictions are made within 3\,$\sigma$ of measurements regarding the extrapolated position. If there is a cluster in the extrapolated layer, the state vector and the covariance matrix is updated with that information. This process is repeated iteratively, in both forward and backward direction in order to smooth the fitted track. Finally, the q/p value, along with azimuthal and zenith angles, is determined on the topmost layer of RPC with a measured point. The reconstructed direction vector is then used to propagate the track back to the top of the building, accounting for all ionization energy losses through the various materials encountered along the way.
Figure \ref{theta_reco_distribution} presents the normalised reconstructed distributions of zenith angle at the top of the building, for both positively charged muons ($\mu^{+}$) and negatively charged muons ($\mu^{-}$) separately. The experimental data is compared with simulated events from three distinct hadronic interaction model combinations: FLUKA-SIBYLL, GEISHA-SIBYLL and UrQMD-SIBYLL, used to generate MC samples in simulation. The selection criteria for the reconstructed zenith angle distribution include a momentum range of 0.8 to 3.0 GeV.

To enhance the reliability of the data by minimizing the impact of mis-reconstruction, additional selection criteria are applied. The first criterion ensures that the fit includes at least 8 layers, implying a minimum path length of $80$\,cm in the mini-ICAL. Additionally, only events with a fit probability greater than $0.015$ are considered.





\subsection{Unfolding}
\label{subsec:unfolding}
The impacts of detector effects can result in events being reconstructed in incorrect momentum/azimuthal angle bins or potentially lost. Both the bin migration and the inefficiency can be corrected  using the unfolding technique.

The response matrix and the background is estimated using the monte carlo simulation. The linear inversion might lead to fluctuations. So regularization techniques have to be applied to get robust results. The TUnfold algorithm\,\cite{Schmitt:2012kp}, interfaced to the ROOT framework, is used to unfold the reconstructed distribution to obtain the true distribution. In this algorithm a least square method of template inversion along with Tikhonov regularization is used\,\cite{Schmitt:2012kp}. 

A low regularization strength, $\tau^2$, results in large fluctuations in the unfolded distribution, whereas high regularization strength biases the results towards the input generator-level distribution.
 The TUnfold algorithm finds an optimum value of regularization strength using an algorithm called L-curve scan. The minimization procedure is described in \cite{Schmitt:2012kp}.


\subsection{Response matrix}

The response matrix is made using reconstructed and simulated objects for different zenith angle ranges for both momentum and azimuthal angle. It will implicitly encode the bin migration, fakes, and inefficiencies. The fakes or background is defined by the events that fall in a particular momentum/azimuthal bin in the region of interest, which is 0.8\,GeV/c to 3\,GeV/c and events simulated will fall outside our region of interest. Also, all the simulated events in the region of interest will not satisfy the selection criteria or the reconstructed value may fall outside the region of interest. So the inefficiency or misses are defined by the events that fall in a particular momentum/azimuthal bin in the region of interest in the Geant4 input level, and reconstructed outside the region of interest or reconstructed in the same bin but did not pass the section criteria or not reconstructed at all. The simulated events from twelve different CORSIKA hadronic model configurations are combined to generate a comprehensive response matrix, which is then used to unfold the data.

\begin{figure}[htbp]
\centering 
\includegraphics[width=1.0\textwidth]{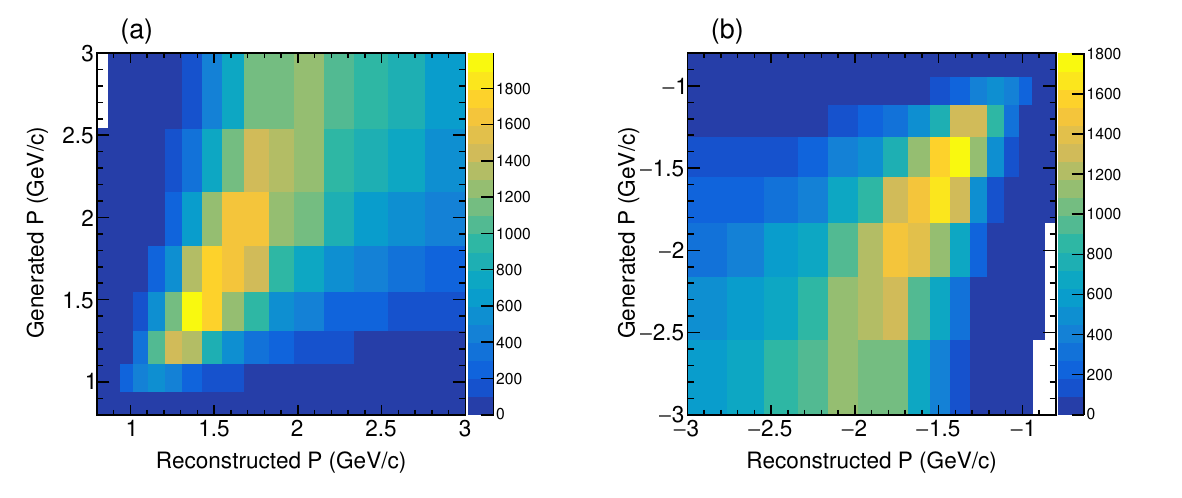}
\caption[The response matrix of momentum for the vertical muon events.]{\label{fig:response_matrix} The momentum response matrix for vertical muon events with zenith angle $\theta$ ranging from $0^\circ$ to $17^\circ$ for, (a) $\mu^+$ and (b) $\mu^-$. The momentum of $\mu^-$ is denoted in negative values. This was done for the ease of distinguishing $\mu^+$ and $\mu^-$ pictorially.}
\end{figure}

Figure~\ref{fig:response_matrix} shows the momentum response matrix, for vertical muon events (i.e., zenith angles $0^\circ$ to $17^\circ$). The integral along the simulated bins adds up to  the efficiency of that particular bin. Due to the short leaver arm and large error in position measurement, the momentum resolution is poor for the high momentum bins and also a saturation in reconstructed momentum, which biases the observed distribution.

\begin{figure}[htbp]
\centering 
\includegraphics[width=1.0\textwidth]{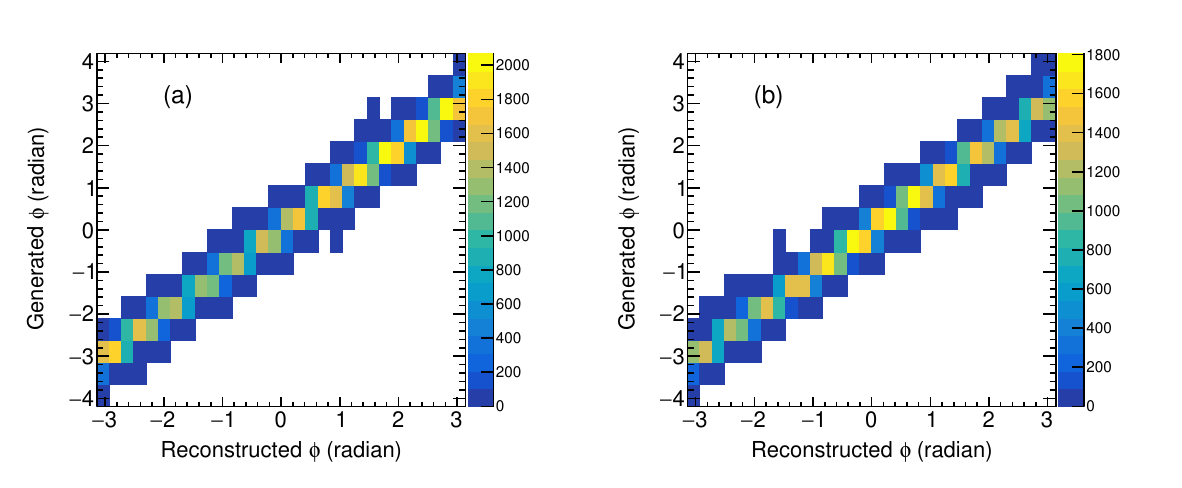}
\caption{\label{fig:response_matrix_phi} The response matrix of azimuthal angle for
  (a) $\mu^+$ and (b) $\mu^-$ events with zenith angle between $17^\circ$ and $26^\circ$.}
\end{figure}

Despite the relatively good resolution of the azimuthal angle, azimuthal angle unfolding is limited to the energy range of 0.8\,GeV/c to 3\,GeV/c and the zenith angle from $17^{\circ}$ - $50^{\circ}$. This restriction is due to the fact that momentum reconstruction is not reliable beyond this range due to large multiple scattering, poor position resolution in comparison with conventional tracking devices and limited lever arm. Moreover, the events that are closer to the zenith (i.e., nearly vertical) tend to have unreliable reconstructed azimuthal angles. To ensure more accurate and meaningful results, we exclude the first theta range from the measurement of the azimuthal angle. At larger zenith angle ($>26^{\circ}$), the momentum range of muons is increased from 0.8\,GeV/c to 0.94\,GeV/c because of very less events at high zenith angles which results in large fraction of fake events.
 The fakes and inefficiency for the azimuthal angle unfolding is estimated in the similar way as it was done for the unfolding of momentum. The response matrix for azimuth angle ($\phi$) is shown in Figure~\ref{fig:response_matrix_phi}. The primary challenge in constructing the unfolding matrix for $\phi$ arises from the fact that $\phi = \pm \pi$ represent the same angle. Consequently, bin migration near $+\pi$ can extend to bins that are distant. To prevent this, the bins at the generator level are widened to encompass a broader angle. The degree of widening depends on the zenith angle and the extent of difference between the reconstructed $\phi$ and the generated $\phi$. If the absolute difference $|\phi_{reco} - \phi_{gen}| \geq \pi$, the generated $\phi_{gen}$ is shifted to a bin that is $2\pi$ away, ensuring proximity to $\phi_{reco}$ and preserving the linear relationship.

\begin{figure}[htbp]
\centering 
\includegraphics[width=0.9\textwidth]{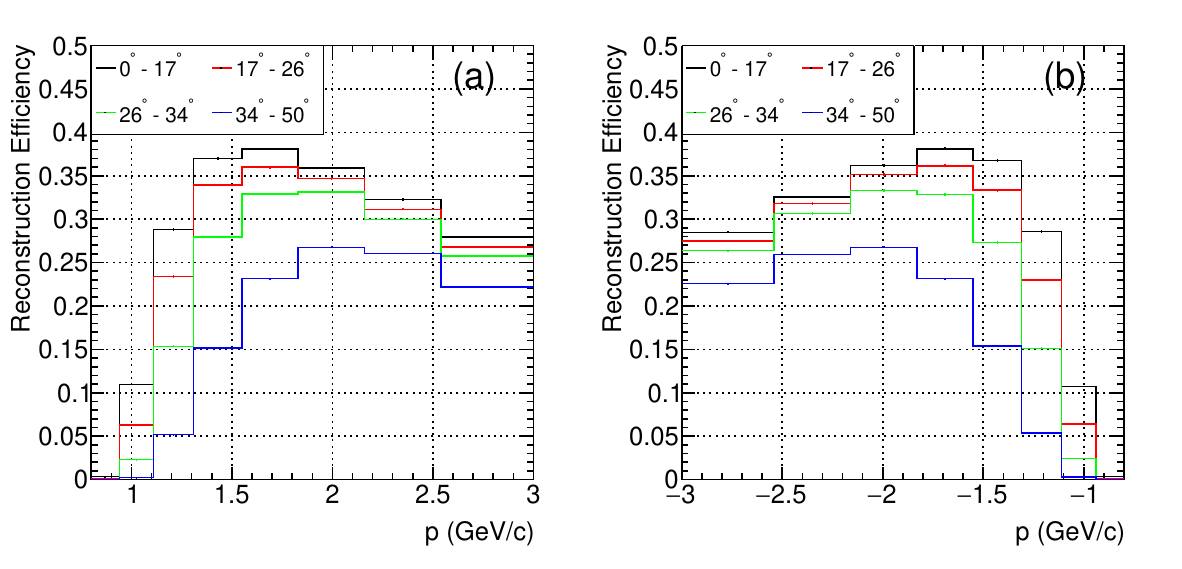}
\caption{\label{fig:reco_efficiency} The reconstruction efficiency for (a) $\mu^+$ (b) $\mu^-$. Here the momentum of $\mu^-$ is denoted in negative values.}
\end{figure}
  
\begin{figure}[htbp]
\centering 
\includegraphics[scale=0.7]{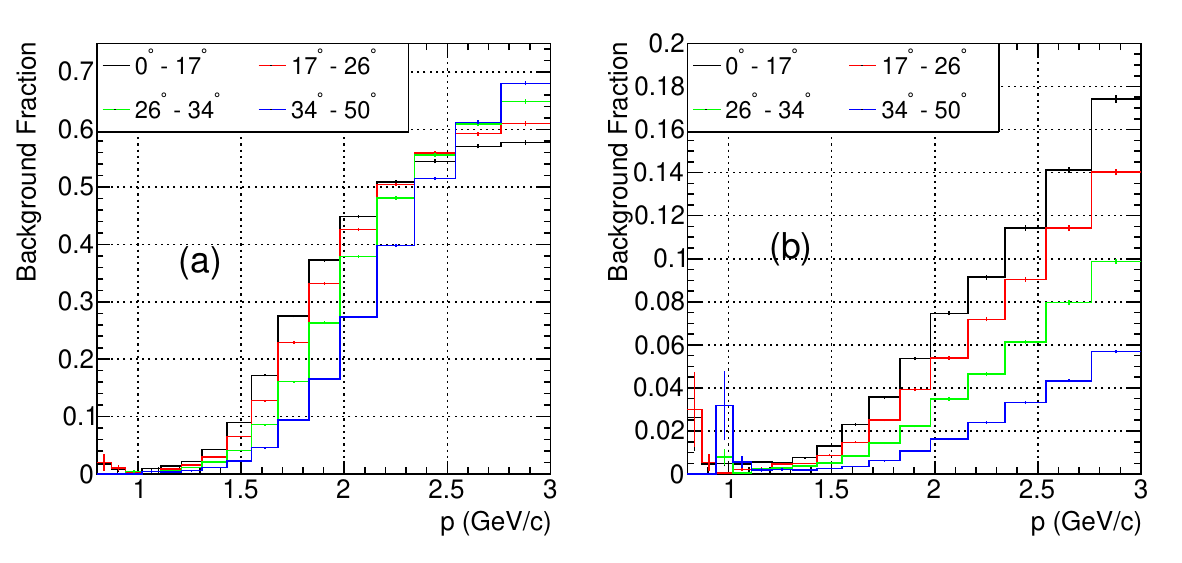}
\caption{The fraction of background events in each reconstructed bin of $\mu^+$ momentum: (a) originating from $\mu^+$, (b) originating from $\mu^-$ (mis-identified).}
\label{fig:reco_fakes_muplus}
\end{figure}

\begin{figure}[htbp]
\centering 
\includegraphics[scale=0.7]{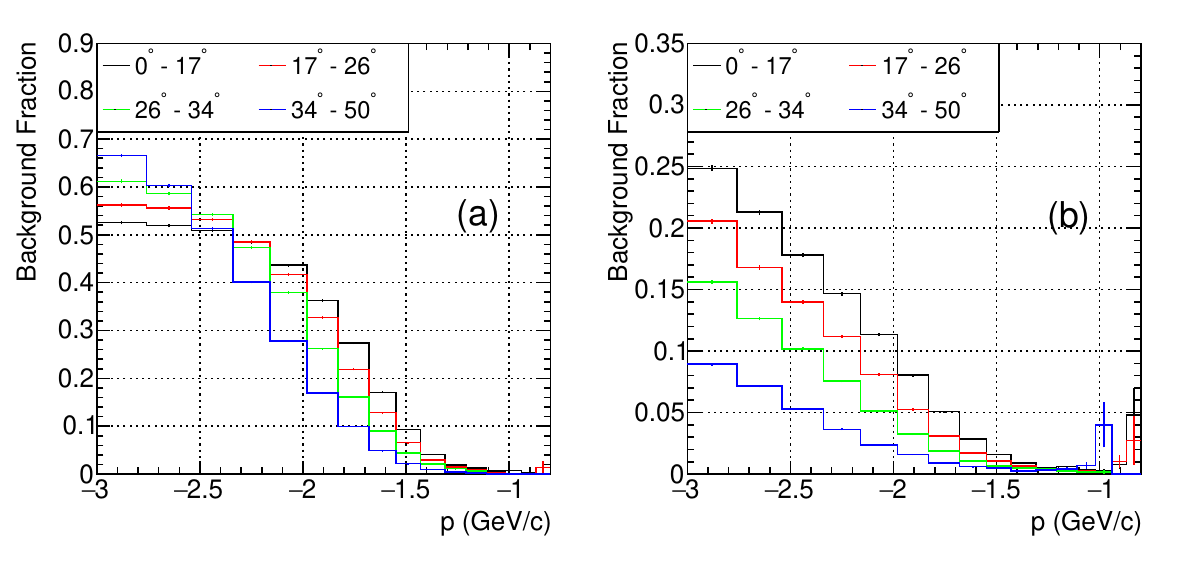}
\caption{The fraction of background events in each reconstructed bin of $\mu^-$ momentum: (a) originating from $\mu^-$, (b) originating from $\mu^+$ (mis-identified). Here, the momentum of $\mu^-$ is denoted with negative values.}
\label{fig:reco_fakes_muminus}
\end{figure}

The reconstruction efficiency within the region of interest is shown in Figure~\ref{fig:reco_efficiency}. The background event fraction in those events is shown in Figure~\ref{fig:reco_fakes_muplus} and Figure~\ref{fig:reco_fakes_muminus} for both $\mu^{+}$ and $\mu^{-}$ respectively. There are two contributions to the background events, (1) charge mis-identification (The $\mu^+$ is identified as $\mu^-$ and vice versa) (2) The reconstructed momentum lying outside the 0.8\,GeV/c to 3\,GeV/c range or migrated within, which is more in comparison to the charge mis-identified fraction.



To validate our unfolding procedure, a closure test is performed using the same Monte Carlo sample employed to create the response matrix. By unfolding this sample and comparing the result to the original input, we confirmed that our response matrix and unfolding process accurately reconstruct the true distribution.





\section{Systematic uncertainities}
\label{sec:systematic_studies}
The uncertainties in the distributions of muon muon momentum and azimuthal angle are related to various factors in simulation and to the selection criteria. Thus the parameters relating to each factor are varied within their uncertainty and the resulting uncertainty in the distributions are evaluated. To minimize the influence of statistical fluctuations, a consistent random number seed is used across all systematic studies, ensuring that event generation remains consistent across all the different scenarios, but the number of simulated events for various systematic studies are much less than the data, consequently the systematic study contains large uncertainties due to the limited number of events.  
The various factors in simulation and the selection criteria are as follows: 
\begin{enumerate}
\item The GEANT4 simulation includes the surrounding walls and roof, but uncertainties exist in the materials and thicknesses. To address this, the roof thickness in the simulation is changed by 10\,$\%$, which is the uncertainty of our estimator.
\item The efficiencies of all the RPCs are increased and decreased by 1$\sigma$, where $\sigma$ is the binomial error on the efficiency for each of the 3\,cm $\times$ 3\,cm pixels \cite{John:2023zqj}.
\item The magnetic field is varied by $3.3\,\%$, the maximum magnetic field inhomogeneity in the central region of the mini-ICAL \cite{Khindri:2023fvf}.
  
\item The uncertainty in the thickness of iron layers is accounted for by simulating with 1\,mm increase and decrease in thickness of the iron layers, which was the tolerance specified during the time of machining. 
  
\item The criteria on the probability of fit $> 0.015$ is applied on reconstructed events. This is varied from 0.012 and 0.018 to estimate the uncertainty due to this selection criteria.
\item The number of layers used in this analysis is 8. This criteria is also varied to 7 and 9, and the difference in those two cases are considered as systematic uncertainty.
\item The variation in the charge ratio due to the use of different CORSIKA models in the unfolding process is assessed. To do this, data is unfolded using three distinct response matrices. Each matrix is generated by combining the Monte Carlo (MC) samples according to their low-energy interaction models:FLUKA, GHEISHA, and UrQMD, for example: the FLUKA response matrix combines all combinations with FLUKA as low energy models (FLUKA-SIBYLL, FLUKA-QGSJET01C, FLUKA-QGSII, FLUKA-VENUS). The unfolded distributions are then compared to the standard distribution, which is derived from a response matrix that combines all twelve models. The average uncertainty across the three distributions is taken as the systematic uncertainty in the comparison.

\end{enumerate}

\begin{figure}[h]
\centering 
\includegraphics[width=1.\textwidth]{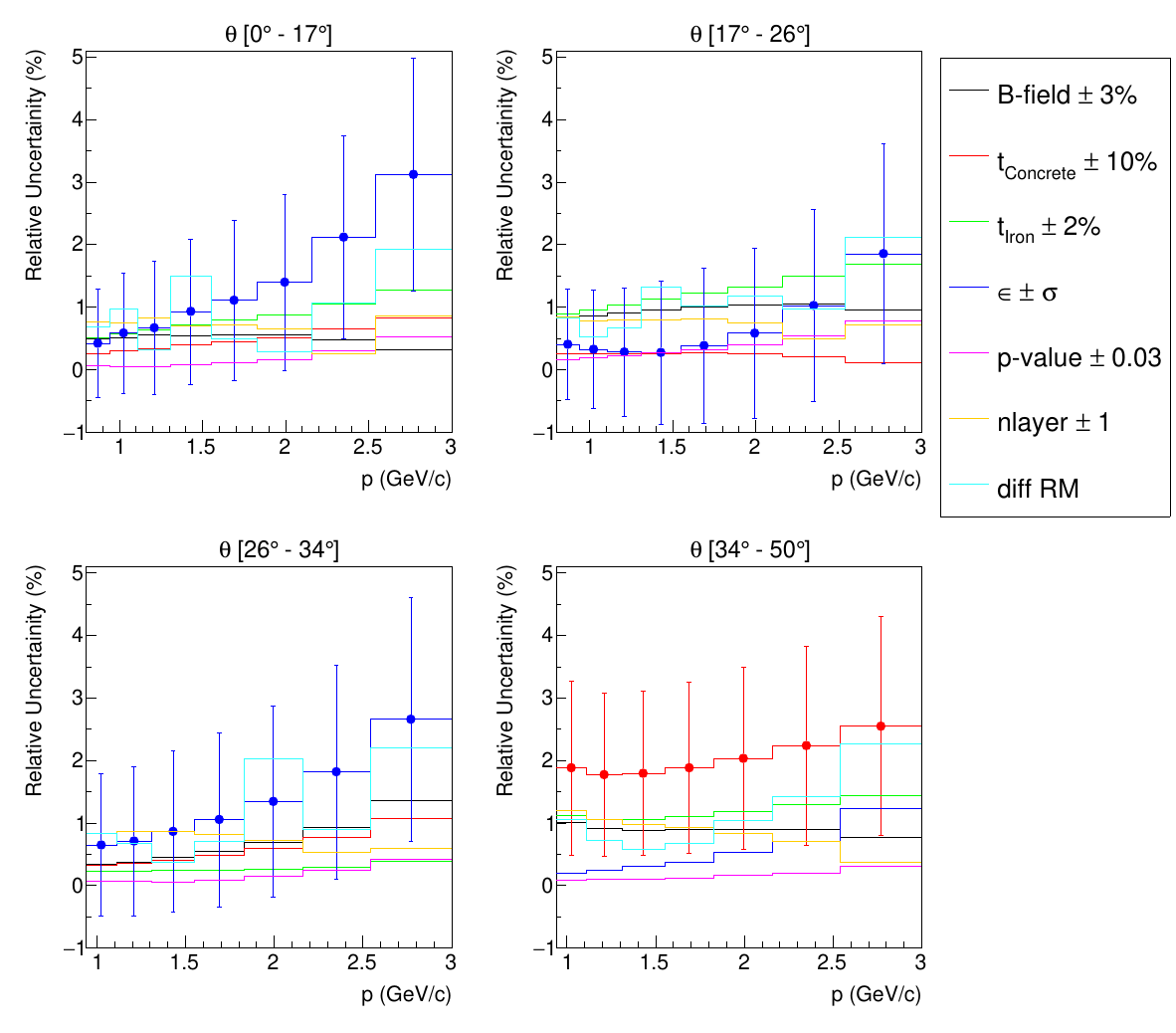}
\caption{The maximum relative uncertainty ($\%$) in each momentum bin resulting from systematic variation in the simulation parameters. All datasets were unfolded using a consistent response matrix derived from the combined events of all models.}
\label{fig:sys1}
\end{figure}


\begin{figure}[h]
\includegraphics[width=1.\textwidth]{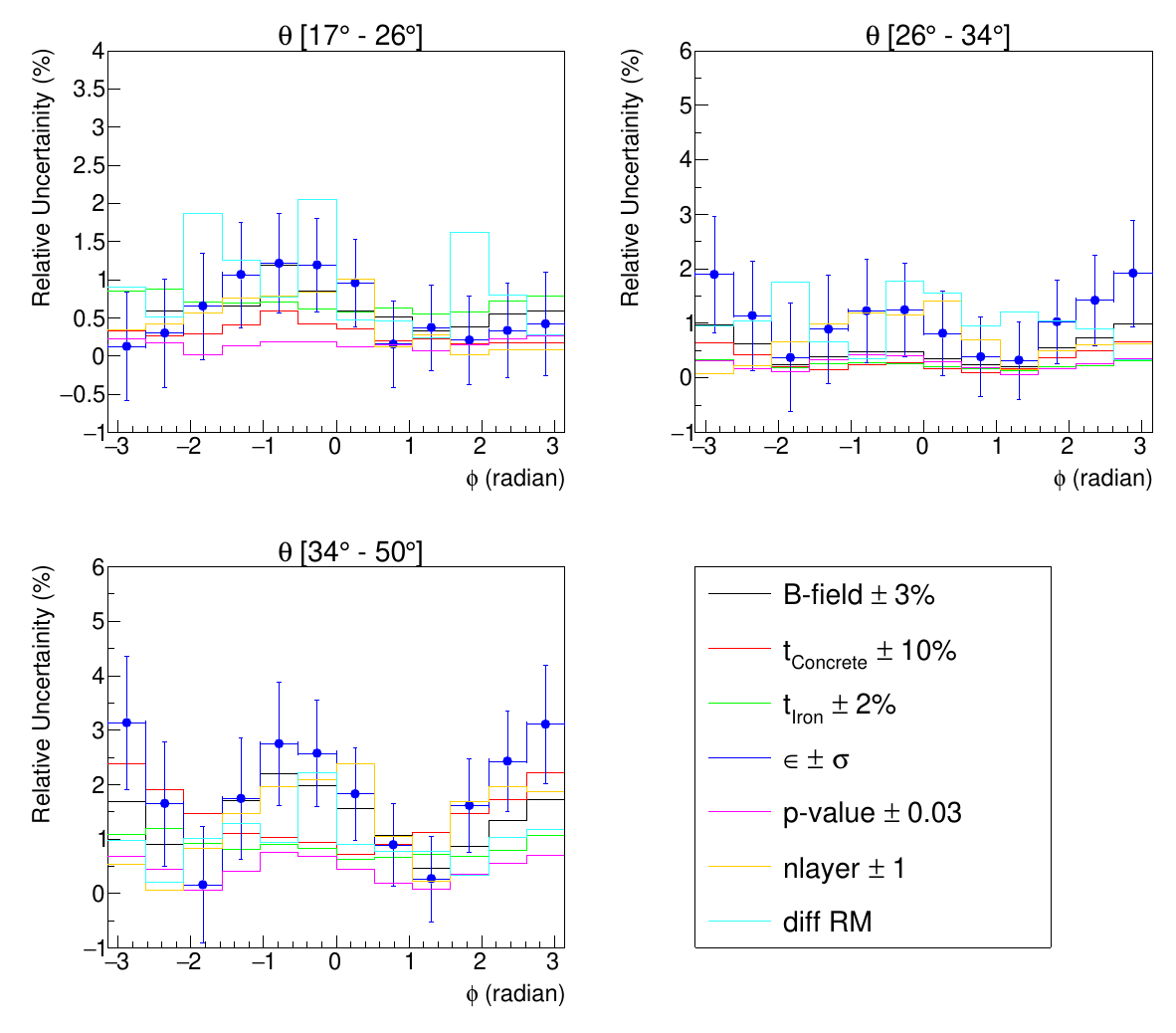}
\caption{The relative uncertainty ($\%$) in the azimuthal angle-dependent charge ratio distribution resulting from systematic variation in the simulation parameters. All datasets were unfolded using a consistent response matrix derived from the combined events of all models.}
\label{fig:sys2}
\end{figure}

The difference between the unfolded distribution obtained using the optimal simulation and those obtained with different systematic variations is used as a measure of uncertainty. The systematic uncertainty is quoted as the maximum deviation observed between the two cases (increase and decrease).
Figure \ref{fig:sys1} and \ref{fig:sys2} shows  maximum relative uncertainty in the unfolded charge ratio distribution for different systematic variation in different zenith angle ranges for both momentum and azimuth angle. The statistical error bar is shown for only one systematic to show the variation is primarily due to the error in unfolding with limited statistics. This is true for other systematics also. In case of momentum, the uncertainty at low zenith angles is mainly dominated by the uncertainty in the strip efficiency while at higher angles it is dominated by concrete-thickness. In all cases, the uncertainties are determined to be less than $4.0\%$. 
 The total systematic uncertainty is calculated by summing the individual uncertainties for each bin in quadrature. These uncertainties are then propagated into the data for each bin.\\
  As a consistency check, unfolding is performed using both the D'Agostini method and the SVD method within the RooUnfold package \cite{Adye:2011gm}, variations of resulting distributions are within the uncertainties of the unfolding error.  Moreover, the GHEISHA-UrQMD datasets were employed to unfold the FLUKA Monte Carlo (MC) dataset. The unfolded distributions were then compared with the original generated values. Similarly, the GHEISHA and UrQMD models were utilized to unfold the FLUKA dataset, as well as in a third combination for cross-validation. In all cases, the observed deviations are within the total uncertainty bounds. There is no variation of unfolded distributions with and without correcting the sample due to mismatch of polar angle as shown in Figure \ref{theta_reco_distribution}.

\section{Results}


\subsection{Momentum dependence of charge ratio of muons at the earth surface}
\label{charge_ratio}
\begin{figure}[h]
\includegraphics[width=\textwidth]{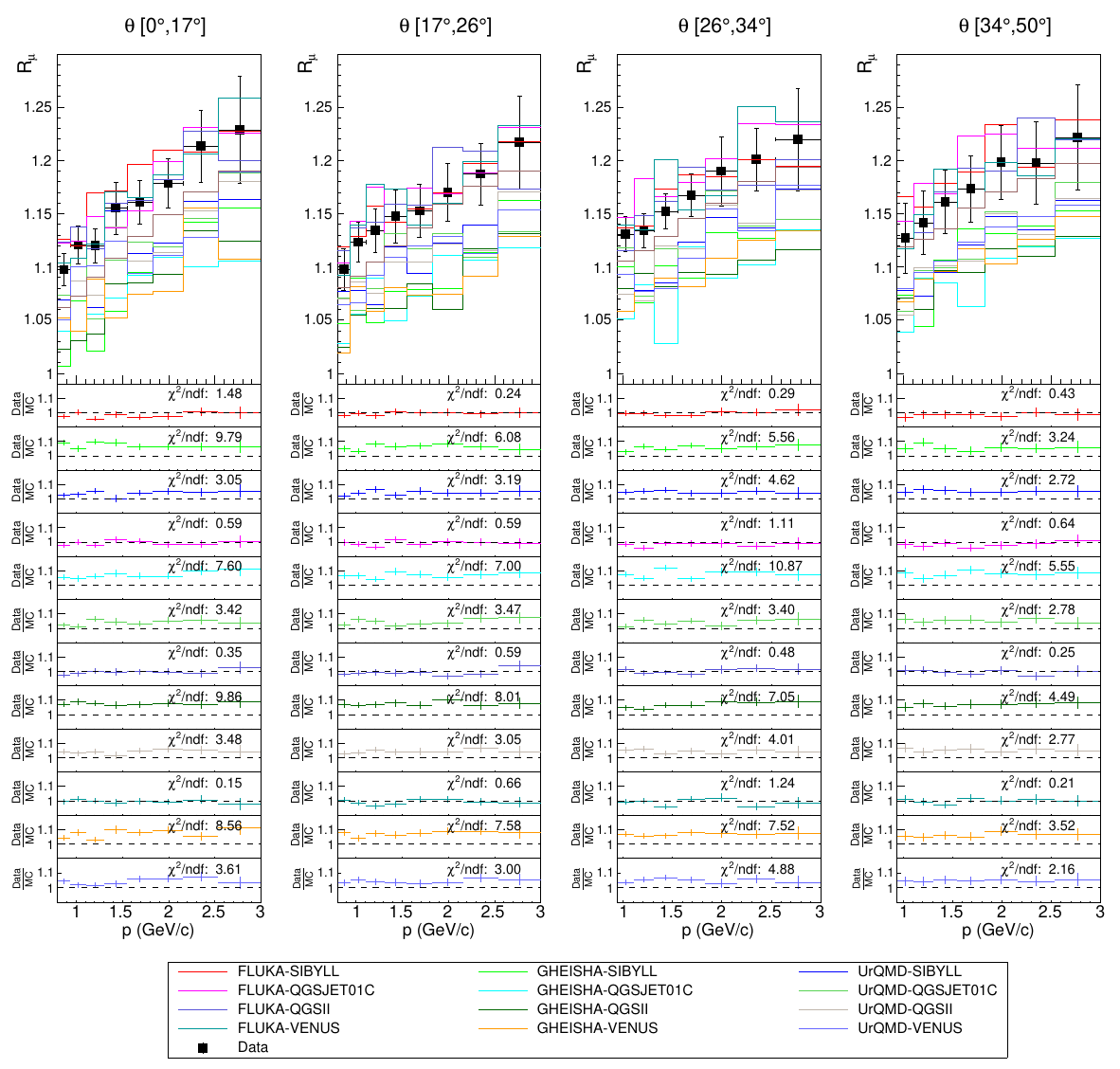}
\caption[The charge ratio in different ranges of zenith angle.]{\label{fig:charge_ratio_mom} The charge ratio of the momentum spectra in different zenith angle ranges. The top row compares the charge ratio obtained from the experiment with the true spectra predicted by various hadronic models in CORSIKA. The bottom rows show the ratio of the experimental data to the prediction of models. The $\chi^{2}/ndf$ represents the reduced chi-squared from these comparisons.}
\end{figure}

The reconstructed output from various different models combined together to constitute the master simulation. This master simulation is utilized to construct the response matrix. Subsequently, each simulation based on different models, as well as the actual data, undergoes unfolding using this response matrix. Before unfolding, the $\mu^{+}$ and $\mu^{-}$ data distribution is scaled by the observed mismatch in the zenith angle distribution between data and MC, as shown in Figure \ref{theta_reco_distribution}. To preserve the shape of the real distribution, both the unfolded and true distributions are scaled by the inverse of the bin width.
 In the investigation of the charge ratio, certain systematic effects associated with the selection criteria are expected to reduce drastically, as they affect both $\mu^+$ and $\mu^-$, e.g., RPC detector efficiency. Consequently, the unfolded results of the charge ratio, $R_\mu = {N_{\mu^+}/}{N_{\mu^-}}$ are deemed a more robust result.  The charge ratio of the unfolded momentum spectra are shown in Figure~\ref{fig:charge_ratio_mom} along with all the MC predictions.
  The study by \cite{Haino:2004nq} indicates that with a higher rigidity cutoff, the charge ratio decreases. At 11\,GV rigidity cutoff, within the same momentum range, their observed charge ratio falls between 1.1 and 1.2 in their study. However, \cite{Haino:2004nq} is primarily observed at lower zenith angles, aligning with our findings in the range of 0$^\circ$ to 17$^\circ$.\\
A distinct separation is observed in the predicted outcomes regarding the charge ratio when utilizing various low-energy models, such as FLUKA and GHEISHA. In contrast, there are minimal discrepancies among the different high-energy models. Thus we combined contributions from four high-energy models: SIBYLL, QGSJET01C, QGSII, and VENUS, for each of the low-energy models: FLUKA, GHEISHA, and UrQMD. The table \ref{tab:chisq_prob_values_mom} lists the resulting $\chi^{2}/ndf$ and fit probability values for the three low-energy models. The charge ratio as a function of momentum in our data exhibits a closer match with FLUKA-simulated models (as shown in figure \ref{fig:charge_ratio_mom} and Table \ref{tab:chisq_prob_values_mom}).

 The charge ratio remains nearly constant up to a zenith angle of $50^\circ$. All the earlier measurements within our momentum range, \cite{beatty2021cosmic} were limited to the zenith angle up to $17^\circ$, which corresponds to our first zenith angle range. To our knowledge, no measurements exist at larger zenith angles for low momentum ranges, making this study a valuable contribution to the hadronic models of cosmic event generators. 



%
%
%
%

\begin{table}[h!]
\centering
\caption{$\chi^{2}/ndf$ and fit probability from various low energy models.}
\label{tab:chisq_prob_values_mom}
\begin{tabular}{|c|c|c|c|c|}
\hline
\multirow{2}{*}{\textbf{Theta Range}} & \multirow{2}{*}{\textbf{Parameter}} & \multicolumn{3}{c|}{\textbf{Low Energy Models}} \\ \cline{3-5}
 & & \textbf{FLUKA} & \textbf{GHEISHA} & \textbf{UrQMD} \\ \hline
\textbf{$0^\circ$- $17^\circ$} & $\chi^{2}/ndf$ & $3.65/8$ & $108.05/8$ & $35.70/8$ \\ 
                 & Probability & $8.87 \times 10^{-1}$ & $9.60 \times 10^{-20}$ & $1.99 \times 10^{-5}$ \\
                 \hline
\textbf{$17^\circ$- $26^\circ$} & $\chi^{2}/ndf$  & $3.00/8$ & $79.51/8$ & $33.53/8$ \\
                 & Probability  & $9.35 \times 10^{-1}$ & $6.14 \times 10^{-14}$ & $4.94 \times 10^{-5}$ \\
                 \hline
\textbf{$26^\circ$- $34^\circ$} & $\chi^{2}/ndf$ & $3.29/7$ & $76.97/7$ & $42.2/7$ \\
                 & Probability & $8.57 \times 10^{-1}$ & $5.71 \times 10^{-14}$ & $4.76 \times 10^{-7}$ \\
                 \hline
\textbf{$34^\circ$- $50^\circ$} & $\chi^{2}/ndf$   & $0.97/7$ & $32.92/7$ & $20.73/7$ \\
                 & Probability & $9.95 \times 10^{-1}$ & $2.74 \times 10^{-5}$ & $4.19 \times 10^{-3}$ \\
                 \hline
                 
\end{tabular}

\end{table}



\subsection{Azimuthal angle dependence of charge ratio of muons at earth surface}
\label{azimuthDistribution}
\begin{figure}[h]
\includegraphics[width=1.0\textwidth]{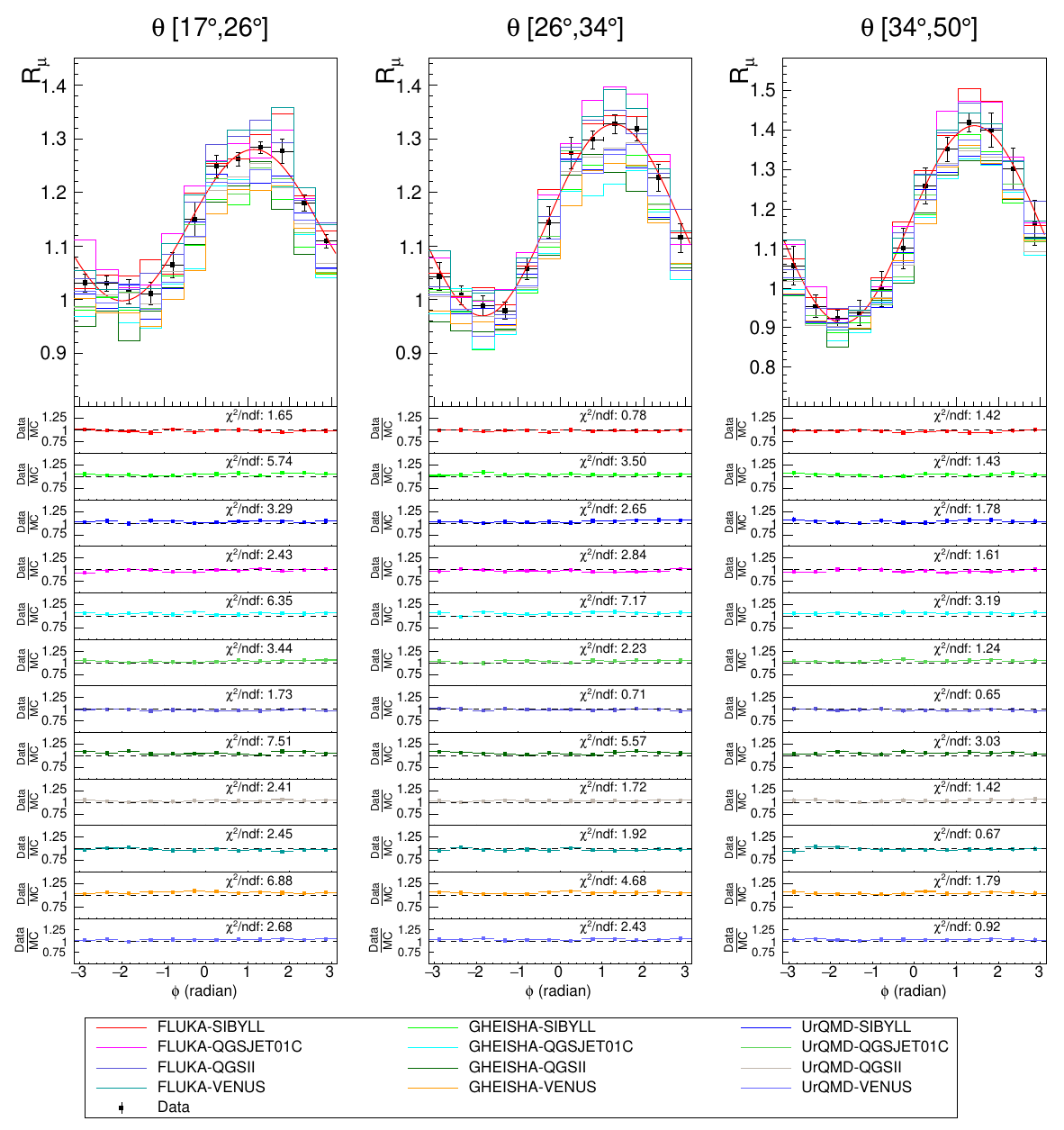}

\caption[The charge ratio in different ranges of zenith angle.]{\label{fig:charge_ratio_phi} The charge ratio of the azimuthal angle spectra in different zenith angle ranges. The top row compares the charge ratio obtained from the experiment (along fit of data with function, \ref{eqn:azimuth}) with the true spectra predicted by various hadronic models in CORSIKA. The bottom rows show the ratio of the experimental data to the predictions from models. The $\chi^{2}/ndf$ represents the reduced chi-squared from these comparisons.}

\end{figure}

The unfolded charge ratio as a function of azimuthal angles, is shown in figure \ref{fig:charge_ratio_phi}.
Specifically, the flux of $\mu^+$ from the west direction, corresponding to an azimuthal angle of $\pi/2$ radians, is higher than that from the east direction ($-\pi/2$), known as the East-West asymmetry of cosmic muons.
Thus, the resulting muon charge ratio exhibits this asymmetry with respect to azimuthal angle.
The generated-level distributions used for unfolding the data are slightly distorted due to contributions from different ranges of zenith angle, as the segregation into these ranges is based on reconstructed zenith angle. Given the uncertainties in the reconstruction of the zenith angle, the reconstructed values may deviate from the generated-level values. This effect is noticed in the case of azimuthal angle, as it is strongly dependent on the zenith angle, but the effect is less than the total uncertainties of the unfolded distributions. Additionally, the impact is particularly significant in the first theta range, which has been excluded from this analysis. There is no visible effect of the smearing of the zenith angle to the ratio in the momentum bin.

 The resulting charge ratio is  found to vary between 0.9 - 1.4. As in the case of momentum, we combine contributions from four high-energy models: SIBYLL, QGSJET01C, QGSII, and VENUS, for each of the low-energy models: FLUKA, GHEISHA, and UrQMD. The table \ref{tab:chisq_prob_values_phi} lists the resulting $\chi^{2}/ndf$ and probability values for the three low-energy models. The distribution of azimuthal angle is matched with FLUKA, which was the case for the distribution of momentum too.

\begin{figure}[h]
\includegraphics[width=1.0\textwidth]{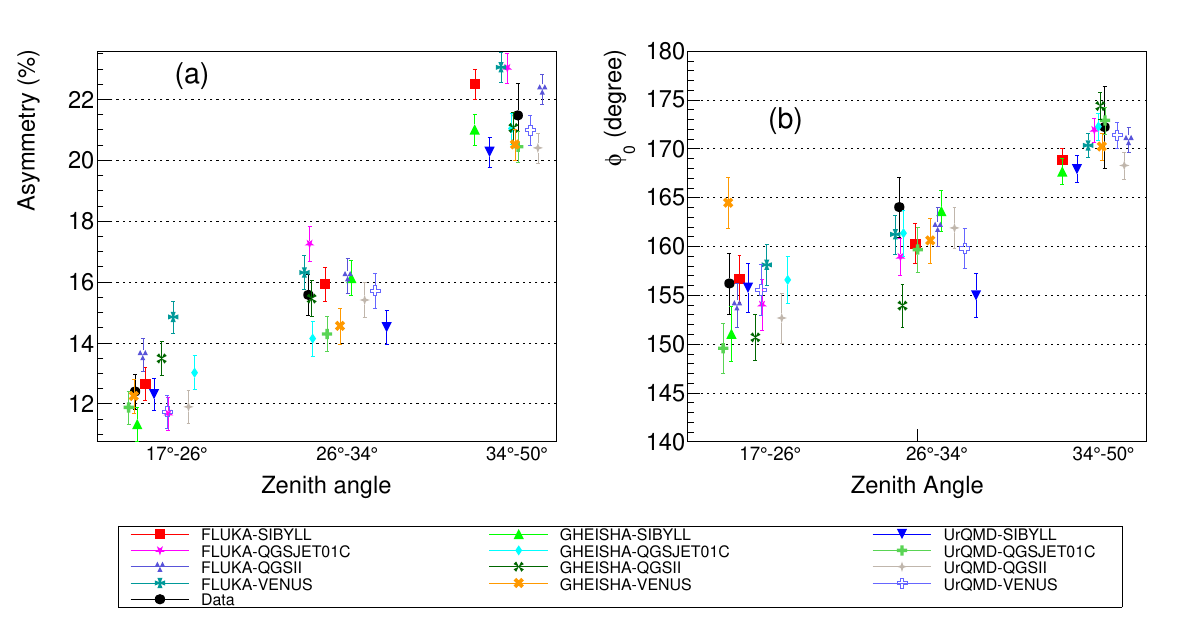}
\caption[Asymmetry in charge ratio.]{\label{fig:charge_ratio_phi_asym}(a) Asymmetry parameter and (b) $\phi_{0}$ in the charge ratio, for data and different model combinations in CORSIKA obtained from the fit. The data points are shifted along the x-axis for better clarity.}

\end{figure}

  The data as well as MC of all the different model combinations are fitted with the function,
\begin{equation}
  f(\phi) = P_{0}(1 + A sin(-\phi + \phi_{0})),
  \label{eqn:azimuth}
\end{equation}
  where the parameters $P_{0}$, A and $\phi_{0}$ are average charge ratio, asymmetry and the phase of the distribution. The asymmetry parameter A and the phase $\phi_{0}$ are shown in figure \ref{fig:charge_ratio_phi_asym}. The observed asymmetry increases with zenith angle, primarily due to the differing rigidity cutoffs experienced by primary cosmic rays entering from the west and east. For example, cosmic rays arriving at a zenith angle of $45^\circ$ experience an energy threshold of 30 GeV when coming from the east, while those entering from the west encounter a lower threshold of 13 GeV (as shown in Figure \ref{fig:rigiditycutoffmadurai}). These values do not show any discriminating power to choose the best model.

%
%
%
%

\begin{table}[h!]
\centering
\caption{$\chi^{2}/ndf$ and fit probability from various low energy models.}
\begin{tabular}{|c|c|c|c|c|}
\hline
\multirow{2}{*}{\textbf{Theta Range}} & \multirow{2}{*}{\textbf{Parameter}} & \multicolumn{3}{c|}{\textbf{Low Energy Models}} \\ \cline{3-5}
 & & \textbf{FLUKA} & \textbf{GHEISHA} & \textbf{UrQMD} \\ \hline
\textbf{$17^\circ$- $26^\circ$} & $\chi^{2}/ndf$ & $20.42/12$ & $119.92/12$ & $55.66/12$ \\
                 & Probability & $5.95 \times 10^{-2}$ & $6.41 \times 10^{-20}$ & $1.38 \times 10^{-7}$ \\
                 \hline
\textbf{$26^\circ$- $34^\circ$} & $\chi^{2}/ndf$  & $18.49/12$ & $83.90/12$ & $34.88/12$ \\
                 & Probability   & $1.02 \times 10^{-1}$ & $7.39 \times 10^{-13}$ & $4.89 \times 10^{-4}$ \\ 
                 \hline
\textbf{$34^\circ$- $50^\circ$} & $\chi^{2}/ndf$ & $10.32/12$ & $31.25/12$ & $18.10/12$ \\
                 & Probability  & $5.88 \times 10^{-1}$ & $1.80 \times 10^{-3}$ & $1.13 \times 10^{-1}$ \\
                 \hline               
\end{tabular}
\label{tab:chisq_prob_values_phi}
\end{table}



\section{Summary}
\label{summary}
This study investigates the momentum and azimuthal angle dependence of the muon charge ratio at the Earth's surface using data obtained from the mini-ICAL detector in the range of 0.8\,GeV/c to 3.0\,GeV/c for a wide range of zenith angle from 0 to 50$^{\circ}$. The charge ratio $R_\mu = N_{\mu^+}/N_{\mu^-}$ is in the range 1.1 to 1.2 and appears to increase with momentum for different zenith angle ranges.
As a function of momentum, the observed charge ratio is in between the predictions of different combinations of CORSIKA models. Also, there is not much variation between different high energy models in CORSIKA. The combination of all high energy models with FLUKA as low energy hadronic model gives better agreement with data. Large deviations are observed with GHEISHA model. Additionally, the ratio remains consistently stable across a range of zenith angles up to $50^\circ$. The asymmetry in charge ratio as a function of azimuthal angle has been measured upto $50^\circ$. It is found to increase with the zenith angle due to varying rigidity cutoffs experienced by primary cosmic rays entering from different directions.

\acknowledgments

We extend our appreciation to the INO collaboration for their indispensable support. We sincerely thank the individuals at IICHEP, Madurai TIFR, Mumbai, and BARC, Mumbai, whose dedicated contributions were integral to the construction and operation of mini-ICAL.



\bibliographystyle{unsrt} 
\bibliography{bibliography.bib}

\end{document}